\documentclass{appolb}
\usepackage{graphicx}
\usepackage{epsfig}
\usepackage{bm}
\usepackage{amssymb}
\usepackage{mathrsfs}
\usepackage{amsmath}
\usepackage{dsfont}
\usepackage{subfigure}

\begin{document}
\title{Flux simulation of the SU(3) spin model at finite chemical potential %
\thanks{Excited QCD - Peniche 6-12 May, 2012.  Presented by Y.Delgado}%
}
\author{ Ydalia Delgado$^{\dagger}$, Christof Gattringer$^{\dagger}$
\address{\small $^{\dagger}$Institut f\"ur Physik, Karl-Franzens Universit\"at, Graz, Austria}
}
\maketitle
\begin{abstract}
We present a Monte Carlo simulation of an effective theory for 
local Polyakov loops at finite temperature and density. 
The sign problem is overcome by mapping the partition sum to a flux representation. 
We determine the phase diagram of the model as a function of the temperature and the chemical potential.
\end{abstract}
\PACS{12.38.Aw, 11.15.Ha, 11.10.Wx}
  
\section{Introduction}
\vspace{-1mm}
\noindent 
Lattice QCD is a powerful tool to
address non-perturbative phenomena quantitatively and in principle is one of the most appropriate techniques
to explore the QCD phase diagram.  However, at finite chemical
potential the fermion determinant becomes complex and it can not be used as a Boltzmann
weight in Monte Carlo simulations.  Alternative approaches such as reweighting, 
power series expansion, strong coupling/large mass expansion or analytic continuation from imaginary chemical 
work only for small chemical potential leaving the rest of the phase diagram unexplored. For true progress
with QCD thermodynamics on the lattice new ideas are necessary.

\vspace*{2mm} \noindent In this article we explore the phase diagram of the SU(3) spin model \cite{wyldkarsch}, 
where the degrees of freedom are traced SU(3) valued spins (local Polyakov loops)  as
 a function of  temperature and chemical potential.
This effective theory can be derived from full QCD using strong coupling expansion for the gluon action
and hopping expansion for the fermion determinant. It is motivated by  the relation
of the deconfinement transition and center symmetry of pure gauge theory \cite{znbreaking}.  
From the fermion determinant one
takes into account a center symmetry breaking term which couples the chemical potential $\mu$
and gives rise to a sign problem at finite $\mu$. However, in this model the sign problem
can be solved by exactly rewriting the partition sum in terms of flux variables \cite{flux,su3}.

\section{Center effective theory}
\vspace{-1mm}
\noindent The action of the center effective theory has the form

\begin{equation}
S \;  = \; - \!\sum_x \left(\! \tau \! \sum_{\nu = 1}^3 \! \Big[ P(x) P(x+\hat{\nu})^\star
+ c.c. \Big] 
+ \kappa \Big[ e^\mu P(x) +  e^{-\mu} P(x)^\star \Big]\! \right)\; .
\label{action_su3} 
\end{equation}

\noindent The degrees of freedom $P(x)$ are the traced
SU(3) variables $P(x) =$ Tr $L(x)$ with $L(x) \in$ SU(3) 
attached to the sites $x$  of a three-dimensional cubic lattice with
periodic boundary conditions. By $\hat{\nu}$ we denote the unit vector in
$\nu$-direction, with $\nu = 1, 2, 3$.  The first term of the action, i.e., the
nearest neighbor interaction term, 
can be obtained as the leading contribution in the strong
coupling expansion of the gauge action. This term is invariant under center
transformations $P(x) \rightarrow z P(x)$ with $z \in \mathds{Z}_3$.  The parameter
$\tau$ depends on the temperature (it increases with $T$) and is real and positive.
The second term, referred to as the magnetic term, is obtained as the leading
contribution in the hopping expansion (large mass expansion) of the fermion determinant.  
The real and positive  parameter $\kappa$ is proportional to
the number of flavors and  depends on the fermion mass (it decreases with $m_q$).
The magnetic term breaks center symmetry explicitly and is complex when the
chemical potential $\mu$ is non-zero, thus generating a sign problem.

\vspace*{2mm}
\noindent The grand canonical partition function of the model described by
(\ref{action_su3}) is obtained by integrating the Boltzmann factor $e^{-S[L]}$
over all configurations of the Polyakov loop variables. The corresponding measure is
a product over the reduced Haar measures $dL(x)$ at the sites $x$. Thus

\begin{equation}
Z \; = \prod_x \int_{SU(3)} dL(x) \, e^{-S[L]} \; = \; \int D[L] \, e^{-S[L]} \; .
\label{sum_su3} 
\end{equation}
Equations (\ref{action_su3}) and (\ref{sum_su3}) define the SU(3) effective
theory.
 
\section{Solving the sign problem}
\vspace{-1mm}
\noindent 
To overcome the sign problem we apply high temperature expansion techniques and 
map the theory onto a flux representation, where the partition function is rewritten
in terms of new degrees of freedom, so called flux variables. Here we outline the
general strategy for the derivation of the flux representation (for the details see \cite{su3}). 
The general steps are:
\vspace{-1mm}
\begin{enumerate} 
\item Write the Boltzmann weight in a factorized form and expand 
the exponentials for individual links and sites.
  \begin{itemize}
  \vspace{-1mm}
  \item Nearest neighbor term (links):
  $$
  e^{\tau P(x)P(x+\hat{\nu})^\star}\ \rightarrow\ 
  \sum_{l_{x,\nu}}\frac{\tau^{l_{x,\nu}}}{l_{x,\nu}!} 
  \big[P(x)P(x+\hat{\nu})^\star\big]^{l_{x,\nu}}\; ; \;
  $$
  $$
  e^{\tau P(x)^\star P(x+\hat{\nu})}\ \rightarrow\ 
  \sum_{\overline{l}_{x,\nu}}\frac{\tau^{\overline{l}_{x,\nu}}}{\overline{l}_{x,\nu}!}
  \big[P(x)^\star P(x+\hat{\nu})\big]^{\overline{l}_{x,\nu}}
  $$
  \item Magnetic term (sites),
  we use $\eta \equiv \kappa e^\mu$ and $\overline{\eta} \equiv \kappa e^{-\mu}$:
  \vspace{-1mm}
  $$
  e^{\eta P(x)}\ \rightarrow\ \sum_{s_x}\frac{\eta^{s_x}}{s_x!} P(x)^{s_x}\;\  ; \;\ 
  e^{\overline{\eta} P(x)^\star}\ \rightarrow\ 
  \sum_{\overline{s}_x}\frac{\overline{\eta}^{\overline{s}_x}}{\overline{s}_x!} P(x)^{\star\ \overline{s}_x}
  $$
  \end{itemize}
  \vspace{-2mm}
\item Rewrite the partition function as:
\begin{equation}
Z =  \sum_{\{l,\overline{l}\}} \sum_{\{s,\overline{s}\}} \!
 \left( \prod_{\overline{x},\nu} 
 \frac{\tau^{l_{x,\nu}+\overline{l}_{x,\nu}}}{l_{x,\nu}!\overline{l}_{x,\nu}!}  \right) \!\!
 \left( \prod_x \frac{\eta^{s_x}\overline{\eta}^{\overline{s}_x}}{s_x!\overline{s}_x!} 
 \int \!\!d P(x) \,  P(x)^{f(x)} P(x)^{\star \, \overline{f}(x)} \right) \!,
\end{equation}
\vspace{-1mm}
\noindent
where $f(x) \; = \; \sum_{\nu=1}^{3} [l_{x,\nu}+\overline{l}_{x-\hat{\nu},\nu}]\ +\ s_x$ and 
$\overline{f}(x) \; = \; 
\sum_{\nu=1}^{3} [l_{x-\hat{\nu},\nu}+\overline{l}_{x,\nu}]\ +\ \overline{s}_x$ 
denote two types of fluxes at a site $x$ of the lattice.
\vspace{-2mm}
\item After integrating out the SU(3) variables $L(x)$ \cite{wipfsu3}, 
the new form of the partition sum depends only on the flux variables:
\vspace{-2mm}
  \begin{itemize}
  \item Dimers  $l_{x,\nu}, \overline{l}_{x,\nu} \in [0,+\infty[$ , living on the links $(x,\nu)$.
\vspace{-1mm}
  \item Monomers  $s_x,\overline{s}_x \in [0,+\infty[$ , living on the sites $x$.
  \end{itemize}
\vspace{-2mm}  
\item The flux variables $l_{x,\nu}, \overline{l}_{x,\nu}, s_x, \overline{s}_x$ are the new
degrees of freedom and $\sum_{\{l,\overline{l}\}} \sum_{\{s,\overline{s}\}}$ denotes the sum
over all their configurations. The flux variables are subject to a constraint which 
forces the total flux $f(x) - \overline{f}(x)$ to be a multiple of 3 at each site $x$.
\end{enumerate}

\section{Numerical analysis}
\vspace{-1mm}
\noindent For the analysis we performed simulations with a local Monte Carlo update on $10^3$, $16^3$ and $20^3$ lattices and 
focused on the bulk observables internal energy $U$ and the magnetization $P$ (which is identified with the Polyakov loop of QCD), as well as their
fluctuations $C$ (heat capacity) and $\chi_P$ (Polyakov loop susceptibility).
 
\begin{figure}[t]
\begin{center}
\includegraphics[width=0.999\textwidth,clip]{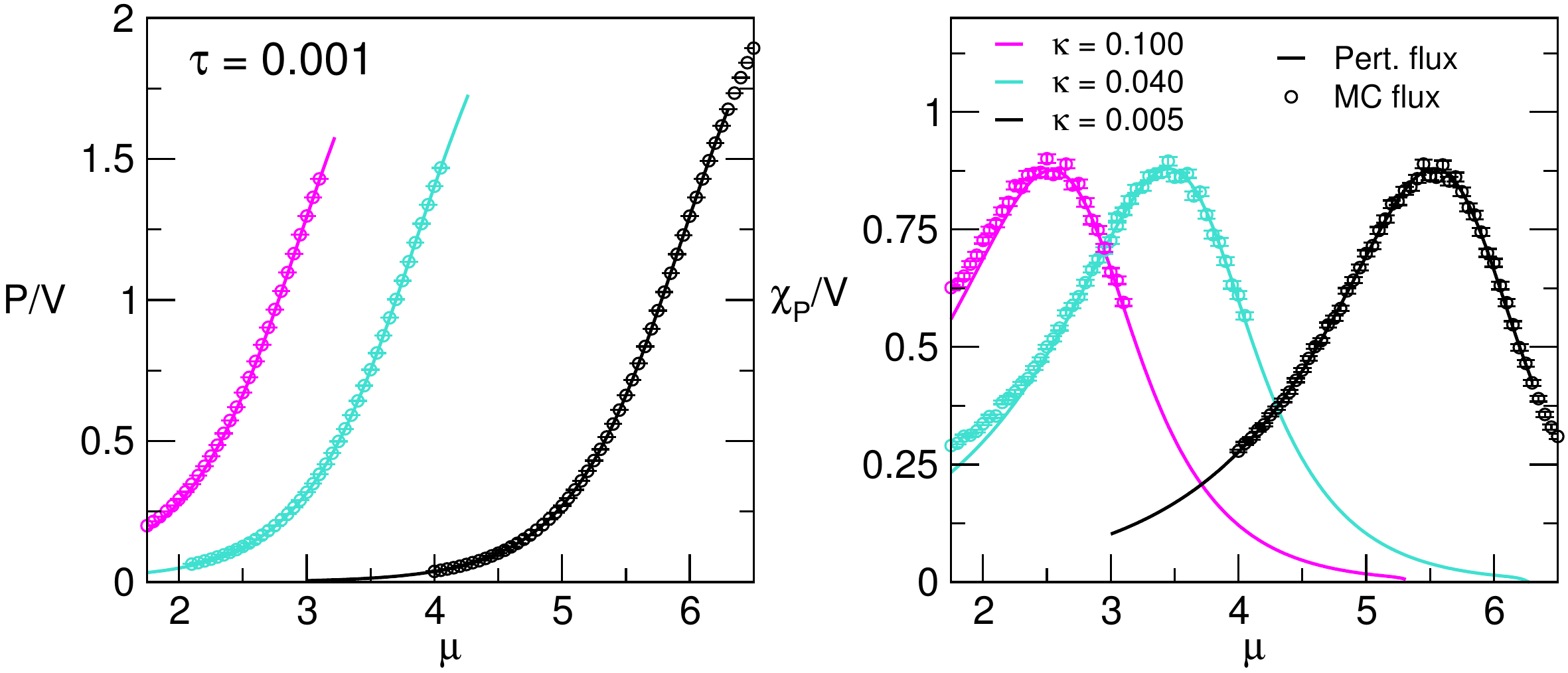}
\end{center}
\vspace{-4mm}
\caption{$P$ (left) and $\chi_P$ (right) for $\kappa = 0.1, 0.04$ and $0.005$ and $\tau = 0.001$. 
We compare the results from the Monte Carlo simulation on a $10^3$ lattice (circles) and 
perturbative expansion in $\tau$ (lines).} \label{mu0}
\end{figure}

\begin{figure}
\centering
\subfigure{
\includegraphics[width=0.474\textwidth,clip]{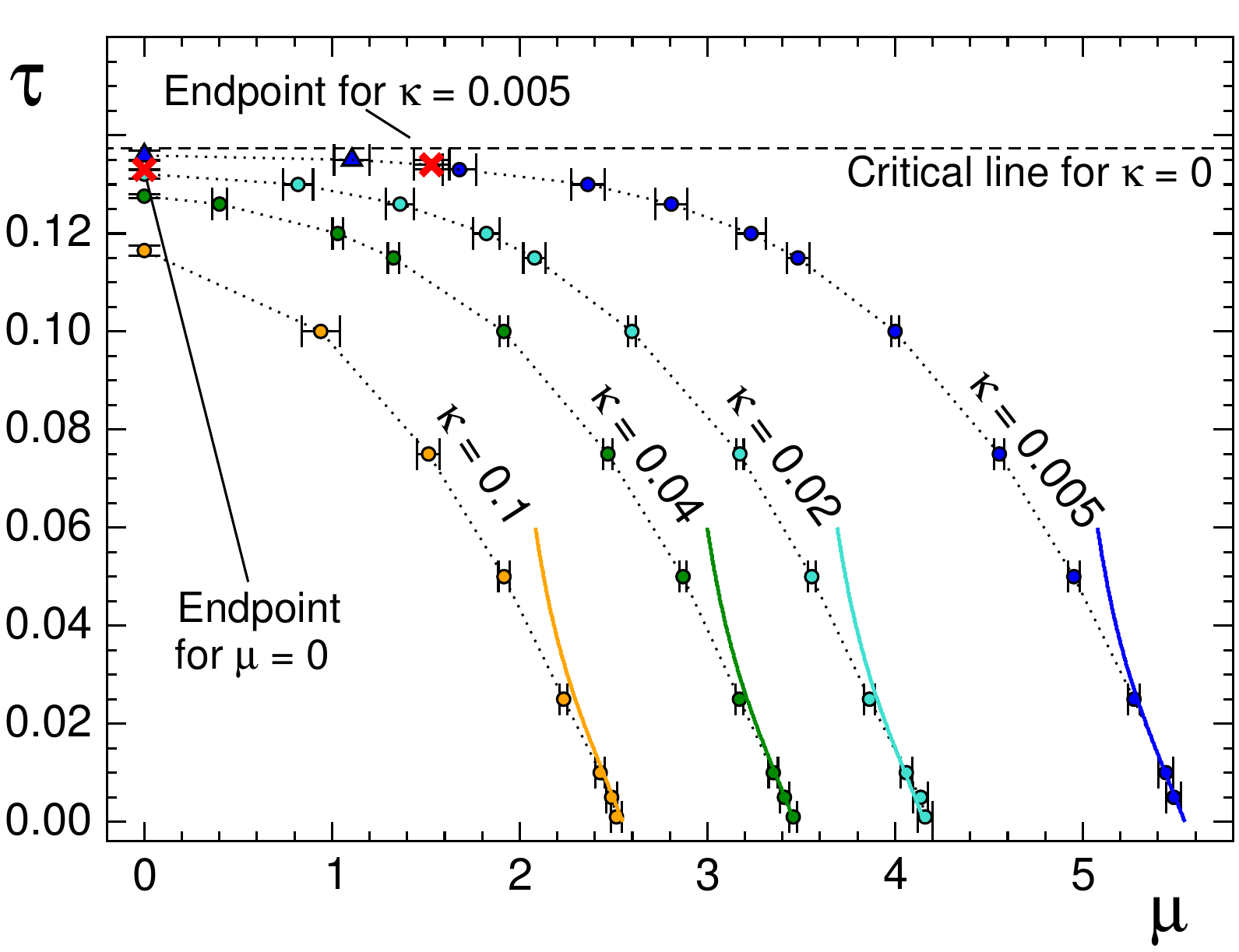}
\label{su3_xp}
}
\subfigure{                       
\includegraphics[width=0.474\textwidth,clip]{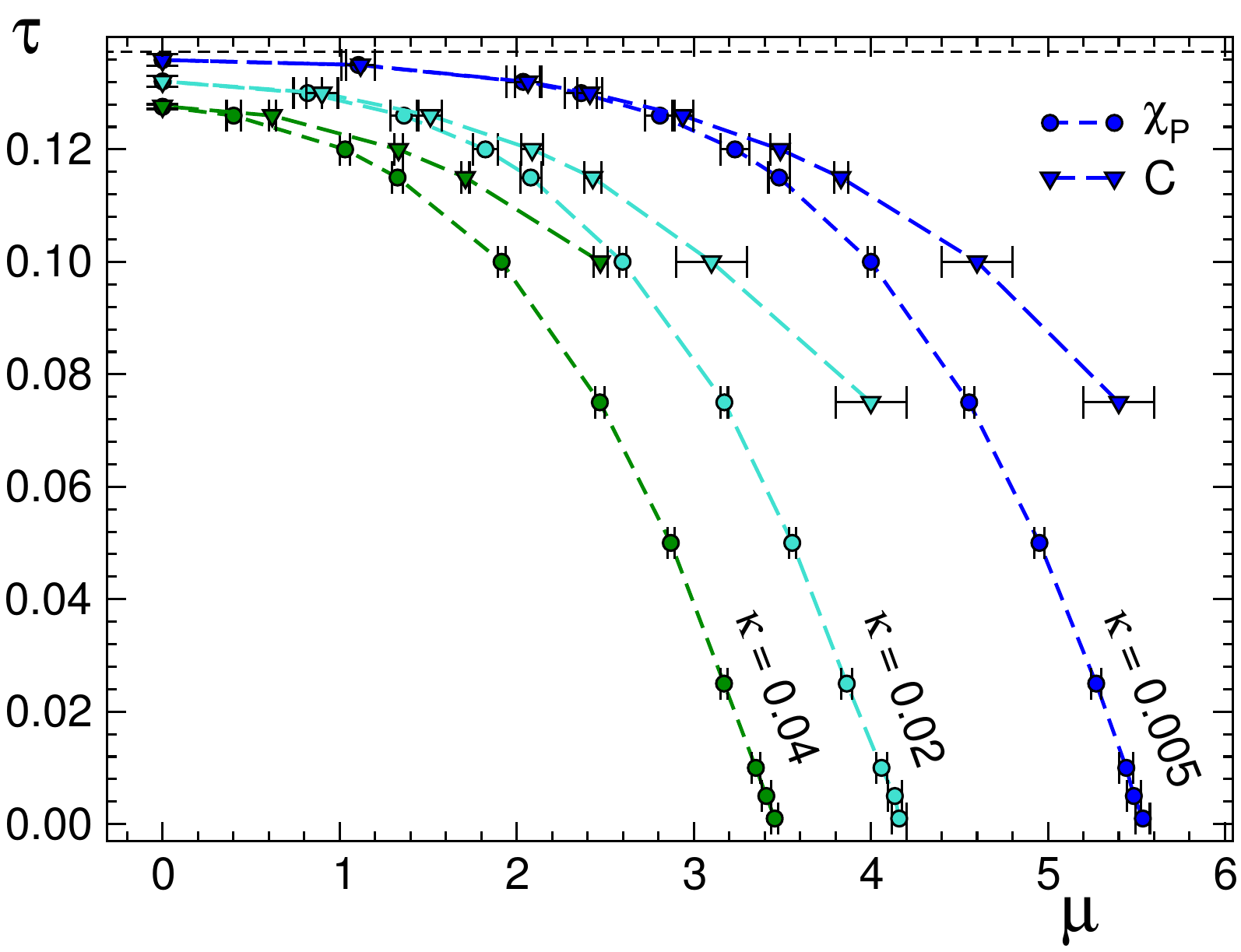}
\label{su3_comparison}
}
\vspace*{-2mm}
\caption{(a) Left: Phase diagram obtained from the maxima of $\chi_P$ for 4 values of $\kappa$.  
The horizontal line marks the critical $\tau$ for $\kappa = 0$, and the curves at the
bottom are the results from a $\tau$ expansion. The red point is the critical end
point for $\kappa = 0$. (b) Right: Comparison of the phase boundaries obtained from
the maxima of $\chi_P$ and $C$ for three values of $\kappa$.}
\end{figure}

\vspace*{2mm} \noindent First we performed several checks of the flux representation and the algorithm.
Fig.~\ref{mu0} shows that for small $\mu$ and $\tau$ the data obtained from the simulation (circles) nicely
approaches the analytical results from a perturbative expansion in $\tau$ (lines). We
plot $P$ and $\chi_P$ for $\tau = 0.001$ and three different values of $\kappa$ as a 
function of the chemical potential.  The same comparison is shown in Fig.~\ref{su3_xp}, 
where the solid curves at the bottom are the positions of the maxima from the perturbative expansion. 

\begin{figure}[h]
\begin{center}
\includegraphics[width=0.85\textwidth,clip]{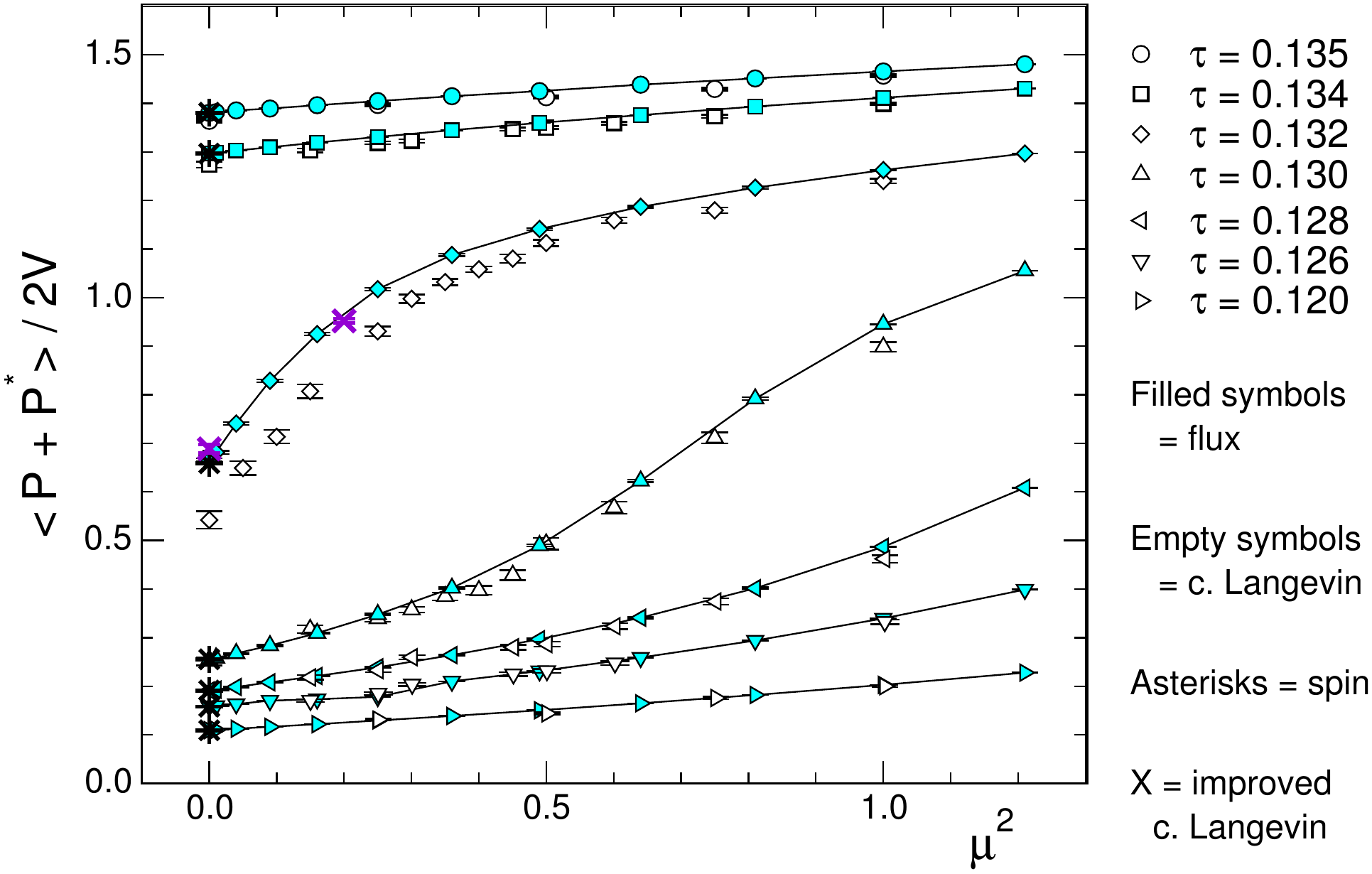}
\end{center}
\vspace{-4mm}
\caption{Comparison of $\langle P + P^\star \rangle/2V$
 from the flux simulation (filled symbols) to the results from the complex Langevin approach 
(empty symbols and two high accuracy data points are marked with crosses). For $\mu = 0$ we also added the results from a simulation in the
conventional spin approach (asterisks). We compare data at different values of $\tau$ as a function of $\mu^2$ 
for $\kappa = 0.02$ on lattices of size $10^3$.}
\label{checksu3}
\end{figure}

\vspace*{2mm} \noindent We also compared our results to other approaches. 
Fig.~\ref{checksu3} shows that the flux results and the data from a complex Langevin calculation  \cite{aarts}
agree very well, and for $\mu = 0$ also with the results
from a conventional simulation in the spin representation. The discrepancy at $\tau = 0.132$ is solved
when a higher-order algorithm is used for the two values, $\mu^2 = 0.0$ and 0.2 (crosses).

\vspace*{2mm} \noindent To explore the phase boundaries in the $\tau$-$\mu$ plane,
we identified the positions of the maxima of $\chi_P$ and $C$.
Subsequently we used two methods to determine the nature of the transitions: 
first we studied the histograms of $U$ and $P$ to search for a
double peak behavior characteristic of a first order transition, and
secondly  we analyzed the volume scaling of $C$ and $\chi_P$.
Fig.~\ref{su3_xp} shows the positions of the maxima of $\chi_P$ in the
$\tau$-$\mu$ plane for $\kappa\ =$ $0.1$, $0.04$, $0.02$ and $0.005$.  
We find that there is a first order phase transition for small
$\mu$ and $\kappa < \kappa_c$ (triangles), while the rest of the transition lines shows a crossover behavior (circles). 
Our estimate for the critical point for $\mu = 0$ is $(\tau_c,\kappa_c) = (0.1331(1),0.016(2))$.
This value is different from  a mean field analysis of the SU(3) spin model \cite{karschMF} 
where the critical point was reported to be at $\kappa = 0.059$.  However, in \cite{splittorff} it
was shown that when considering higher order corrections the value $\kappa_c$ from the 
mean field approach decreases.
Fig.~\ref{su3_comparison} shows the positions of the maxima of $\chi_P$ and $C$,
demonstrating that the crossover region (manifest also in different positions 
for the maxima of $\chi_P$ and $C$) becomes wider with increasing $\mu$.

\section{Conclusions}
\vspace{-1mm}
\noindent We have studied an effective theory for the Polyakov loop at finite temperature and density.  
Mapping the theory onto a flux representation enables us
to have a model free of the sign problem and opens the possibility to use Monte Carlo techniques.
For large values of $\kappa$ (physical case) the transition is of a smooth crossover 
type and we conclude that center symmetry alone does not provide a mechanism for first
order behavior in the QCD phase diagram.

We also compared the recently published results from a complex Langevin simulation of the SU(3)
spin model \cite{aarts} to the data from our flux simulation.
We find very good agreement between the two methods which is a valuable test for
both, the flux and the complex Langevin approach.

\section*{Acknowledgments} 
\vspace{-1mm}
\noindent
We thank Gert Aarts, Hans Gerd Evertz, Daniel G\"oschl, Frank James and
Christian Lang for fruitful discussions at various stages of this work, and
Gert Aarts and Frank James also for providing the complex Langevin data. This work was supported 
by the Austrian Science Fund, FWF, DK {\it Hadrons in Vacuum, Nuclei, and Stars} 
(FWF DK W1203-N16)
and by the Research Executive Agency (REA) of the European Union 
under Grant Agreement number PITN-GA-2009-238353 (ITN STRONGnet).

\end{document}